\documentclass[journal,twoside,web]{ieeecolor}
\usepackage{generic}
\usepackage{cite}
\usepackage{siunitx}
\usepackage{amsmath,amssymb,amsfonts}
\usepackage{verbatim}

\usepackage{upgreek}
\usepackage{algorithmic}
\usepackage{textcomp}
\DeclareMathOperator\atanh{atanh}
\usepackage{graphicx}
\usepackage{listings}
\usepackage{xcolor}

\definecolor{codegreen}{rgb}{0,0.6,0}
\definecolor{codegray}{rgb}{0.5,0.5,0.5}
\definecolor{codepurple}{rgb}{0.58,0,0.82}
\definecolor{backcolour}{rgb}{0.88,0.96,1} 
\lstdefinestyle{mystyle}{
	backgroundcolor=\color{backcolour},   
	commentstyle=\color{codegreen},
	keywordstyle=\color{magenta},
	numberstyle=\tiny\color{codegray},
	stringstyle=\color{codepurple},
	basicstyle=\sffamily\scriptsize,
	breakatwhitespace=false,         
	breaklines=true,         
	frame = none,   
	captionpos=b,                    
	keepspaces=true,                 
	numbers=left,                    
	numbersep=5pt,                
	showspaces=false,   
	prebreak=\mbox{{\color{gray}\tiny$\searrow$}},             
	showstringspaces=false,
	showtabs=false,                  
	tabsize=2
}

\usepackage{url}
\makeatletter
\let\NAT@parse\undefined
\makeatother
\usepackage{hyperref}
\hypersetup{
	colorlinks=true,
	linkcolor=blue,  
	urlcolor=blue,
	citecolor=blue
}
\def\BibTeX{{\rm B\kern-.05em{\sc i\kern-.025em b}\kern-.08em
    T\kern-.1667em\lower.7ex\hbox{E}\kern-.125emX}}
\usepackage{orcidlink}
\begin{document}
\title{Robust Simulation of Poisson's Equation in a P-N Diode Down to 1\,$\upmu$K}
\author{\vspace{0.5cm}Arnout Beckers\,\orcidlink{0000-0003-3663-0824} \thanks{A. Beckers is with imec, Kapeldreef 75, 3001 Leuven, Belgium.}\thanks{(arnout.beckers@imec.be)}}

\maketitle

\begin{abstract}
Semiconductor devices are notoriously difficult to simulate at deep-cryogenic temperatures. The lowest temperature that can be simulated today in commercial TCAD is around 4.2\,K, possibly 100\,mK, while most experimental quantum science is performed at 10\,mK or lower. Besides the challenges in transport solvers, one of the main bottlenecks is the non-convergence in the electrostatics due to the extreme sensitivity to small variations in the potential. This article proposes to reformulate Poisson's equation to take out this extreme sensitivity and improve convergence. We solve the reformulated Poisson equation for a p-n diode using an iterative Newton-Raphson scheme, demonstrating convergence for the first time down to a record low temperature of one microkelvin using the standard IEEE-754 arithmetic with double precision. We plot the potential diagrams and resolve the rapid variation of the carrier densities near the edges of the depletion layer. The main Python functions are presented in the Appendix.
\end{abstract}


\section{Introduction}
\label{sec:introduction}
\IEEEPARstart{C}{ryogenic} temperatures have been posing challenges to device simulators since the 1980s \cite{selberherr,kantner,gao,mohiyaddin_multiphysics_2019}. The lowest simulate-able temperatures have improved today, but the problem remains essentially the same. The tail of the Fermi-Dirac (FD) distribution becomes almost abrupt at these temperatures, leading to (i) underflow, (ii) bad convergence, and (iii) sharp density variations requiring ever finer meshing.

Variable precision arithmetic is sometimes used to combat these issues, but it is not preferred given the runtime penalties and lack of support for exotic number formats \cite{richey,tedpaper,zlatan}. More than octuple precision (256-bit) is required to simulate devices at \SI{10}{\milli\kelvin}, e.g., for quantum applications. Furthermore, the abrupt 0-K approximation of the FD function has been applied \cite{catapano,aouad}, but it cannot be used if electrothermal differences between, say, \SI{4.2}{\kelvin} and \SI{10}{\milli\kelvin} are to be resolved. 

Recently, commercial TCAD vendors have started dedicated efforts to simulating semiconductor devices at deep-cryogenic temperatures \cite{zlatan,beaudoin_robust_2022,moroz}. In one TCAD tool, specialized in quantum dots for qubits, it is possible to converge at $\SI{100}{\milli\kelvin}$ using a unique adaptive meshing strategy \cite{beaudoin_robust_2022}. In another tool, transport at \SI{4}{\kelvin} is possible using a new Quasi-Fermi Transport solver that prevents numerical cancellation of small drift and diffusion currents \cite{zlatan}. These are important leaps forward, but, to reach lower temperatures, more efforts are needed. 

Here, we take a different approach, focusing on the one-dimensional $p\textendash n$ diode as a toy problem for the electrostatics. We present a minimal working example at \SI{1}{\micro\kelvin} written in Python code, to illustrate and sort out the numerical issues. The impact of various cryo-phenomena is omitted (e.g., dopant freeze-out, band tails, Boltzmann vs. FD statistics, etc.), which can be added later in a robust numerical solver. Section \ref{sec:structure} describes the details of the $p\textendash n$ diode. Section \ref{sec:challenges} discusses the numerical challenges in Poisson's equation. Section \ref{sec:transf} reformulates Poisson's equation to improve convergence. Section \ref{sec:appli3} compares the performance of two different Poisson solvers: \begin{itemize}
	\item \verb|solve_poisson_standard| solves the regular Poisson equation in double precision arithmetic;\\ $\Longrightarrow$ fails to converge below \SI{10}{\kelvin} (reference case)
	\item \verb|solve_poisson_reform| implements the reformulated Poisson equation in double precision; \\ $\Longrightarrow$ converges down to \SI{1}{\micro\kelvin}, possibly at lower temperatures, but this is currently sufficient for the typical quantum applications of today ($\approx$\,\SI{10}{\milli\kelvin}), and also prepares for the upcoming microkelvin science \cite{microkelvin,samani_microkelvin_2022}. 
\end{itemize} 

\section{\label{sec:structure}Specific Background \& Details of the Simulated Structure}
The low-temperature $p\textendash n$ junction is a fundamental part of cryo-CMOS devices and on-wafer test structures for the fabrication of silicon qubits \cite{kao}. It is also an important device in its own right, e.g., $p\textendash n$ diodes were recently experimentally studied for their functioning as temperature sensors in quantum control ICs \cite{hart}. On the other hand, the internal diode variables like charges and potentials, remain largely unexplored at these temperatures due to the lagging simulation support. 

The state-of-the-art simulate-able temperature of the electrostatics in a diode using the standard double precision arithmetic is currently at \SI{30}{\kelvin} (see Fig.2 in \cite{kantner}). In this work, we will reduce this temperature with seven orders of magnitude while staying within double precision. To be precise, the diode simulated in \cite{kantner} is actually a $p\textendash i\textendash n$ diode, but both diodes have sharp internal diffusion layers which are numerically problematic and thus both can serve as good examples. 

\begin{figure}[t]
	\centering
	\includegraphics[width=0.43\textwidth]{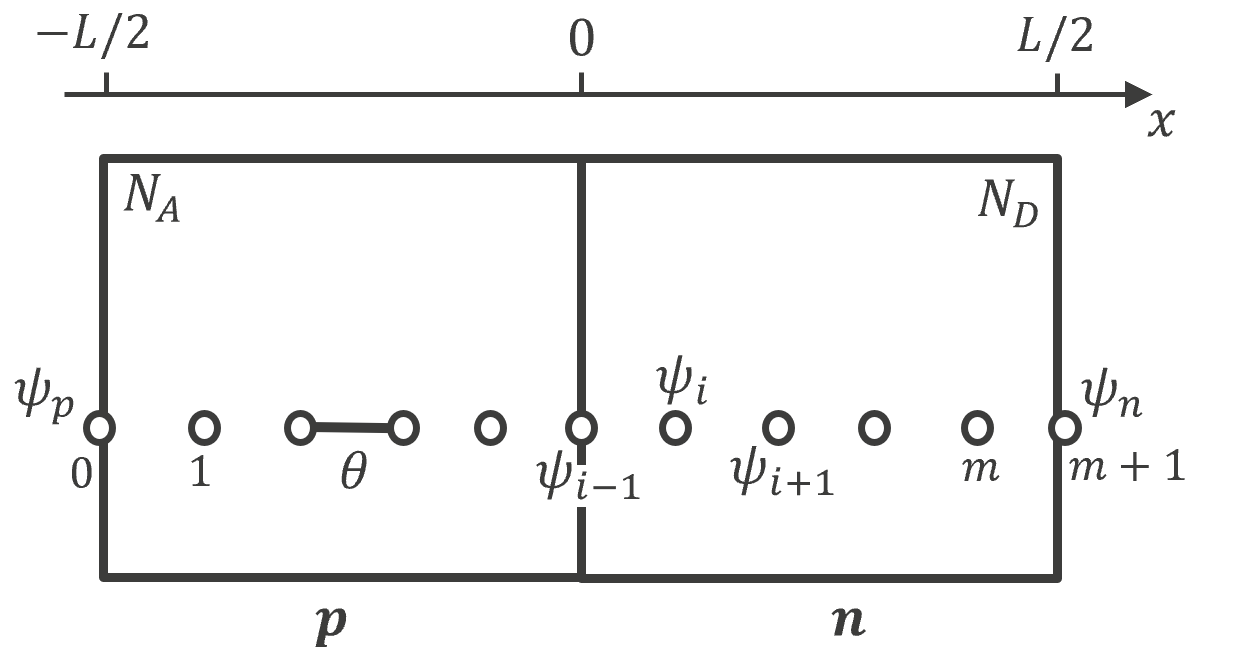}
	\caption{Details of the silicon $p\textendash n$ diode (in 1-D) and discretization.}
	\label{fig:detailspn}
\end{figure}

Fig. \ref{fig:detailspn} shows the details of the $p\textendash n$ diode to be simulated. The basic device is assumed to be made from silicon; non-degenerately doped, fully ionized, and in thermal equilibrium. The acceptor doping concentration on the left side of the junction is $N_{A}=10^{16}\,\si{\per\centi\meter\cubed}$. The donor doping concentration on the right side is $N_D=10^{16}\,\si{\per\centi\meter\cubed}$. The length of the diode is set to $L=\SI{1.2}{\micro\meter}$. The grid size is then $\theta=L/m$, where $m$ is the number of discretization points inside the device. Charge neutrality is imposed at the boundaries on both sides of the junction (i.e., Neumann boundary conditions $d\psi/dx=0$), therefore $\psi_p=\psi_0=\psi_1$ and $\psi_m=\psi_{m+1}=\psi_n$), where $\psi_p$ and $\psi_n$ are the known boundary potentials. 
\section{\label{sec:challenges}Challenges in Poisson's Equation}
Poisson's equation for the diode in Fig. \ref{fig:detailspn} reads
\begin{equation}
	\frac{\partial^2 \psi}{\partial x^2}=\frac{-\rho(\psi)}{\varepsilon_{si}},
	\label{eq:poisson}
\end{equation}
where 
\begin{equation}
	\rho(\psi)=q\cdot \left(-n+p-N_A+N_D\right),
\end{equation} 
is the charge density, $q$ the electric charge, $n$ the electron density, $p$ the hole density, and $\varepsilon_{si}$ the permittivity of Si. The electrostatic potential is defined as $\psi\triangleq -(E_i-E_F)/q$, where $E_i$ is the intrinsic energy, and $E_F$ is the Fermi level. 

\begin{figure}[t]
	\centering
	\includegraphics[width=0.35\textwidth]{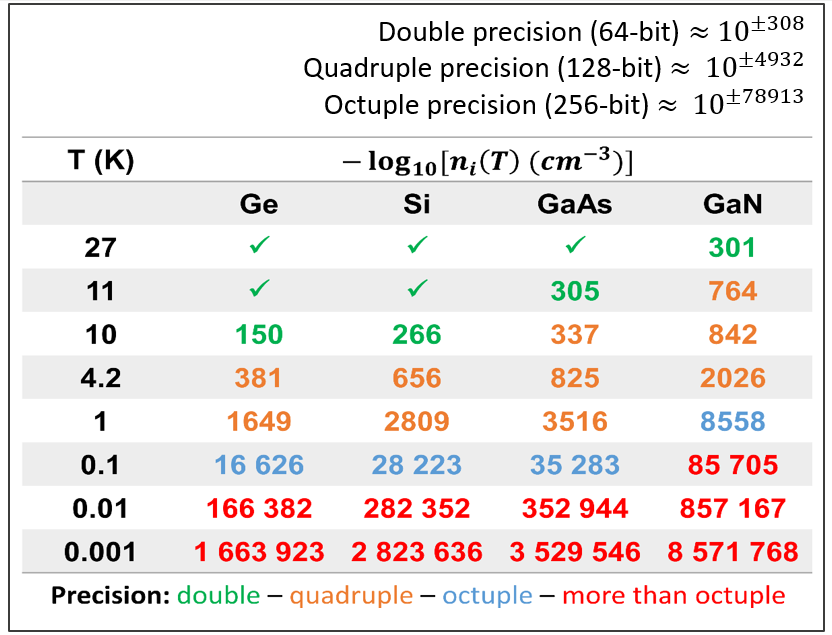}
	\caption{Intrinsic carrier concentration ($n_i$) versus temperature ($T$) for different materials, see (\ref{eq:ni}). Extensions to quadruple or octuple precision formats do not suffice for reaching the lowest temperatures.}
	\label{fig:ni}
\end{figure}

Focusing purely on the numerical challenges, the Boltzmann approximation for $n$ and $p$ provides a good test case for our present purposes, because Boltzmann's exponential tail can both underflow above $E_F$ and overflow below $E_F$. This makes it numerically more challenging than the FD distribution, which can only underflow above $E_F$. The Boltzmann relation for the electron density is given by 
\begin{equation}
	n=n_i\cdot\exp\left(\frac{q\cdot \psi}{k_BT}\right),
	\label{eq:n}
\end{equation}
which brings a three-fold numerical challenge to Poisson's equation at deep-cryogenic temperatures: 
\begin{itemize}
	\item \textbf{(i)} the intrinsic carrier concentration, $n_i$, can easily underflow at these temperatures, which is often mentioned in the literature \cite{tedpaper,kantner}. The table in Fig. \ref{fig:ni} shows the extremely low values that $n_i$ takes below \SI{10}{\kelvin}, which were computed using variable precision arithmetic. These numbers are of limited practical value.
	\item \textbf{(ii)} Since $\psi$ can range up to a few \si{volt}s during operation, and $k_BT/q$ can go below $\sim$\si{\milli\volt}, the used floating-point format cannot accommodate such large exponents in (\ref{eq:n}), causing arithmetic underflow and overflow when numbers fall outside the range of a given precision format ($\approx 10^{-a}<x<10^{a}$), where $a=308$ for standard IEEE-754 double precision arithmetic. The exponential factor can both underflow and overflow depending on the sign of $\psi$ during device operation. 
	\item \textbf{(iii)} $n$ is extremely sensitive to small fluctuations in $\psi$, due to the small $k_BT$ in the exponent of (\ref{eq:n}).
\end{itemize} 
The same remarks also apply to $p$.
\section{\label{sec:transf}Transformations in Poisson's Equation to Improve Numerics and Convergence}
Step-by-step transformations are applied to (\ref{eq:poisson})-(\ref{eq:n}) to improve numerical robustness and convergence at low $T$. 
\subsubsection{Avoid Using the Intrinsic Carrier Concentration} While the values of $n_i$ given in the table in Fig. \ref{fig:ni} are numerically correct, they are far from being physically meaningful in realistically sized semiconductor devices. Using $n_i=10^{-656}$ \si{\per\cubic\centi\meter} is an attempt to model one thermally generated electron in a piece of intrinsic Si with a volume of $10^{650}$ \si{\meter\cubed} at \SI{4.2}{\kelvin}. Thus, it is recommended to avoid using $n_i$ and expand it in (\ref{eq:n}), i.e.,
\begin{equation}
	n_i=\sqrt{N_cN_v}\cdot \exp\left(\frac{-E_g}{2k_BT}\right),
	\label{eq:ni}
\end{equation} 
which gives the safer expression,
\begin{equation}
	n=N_c\cdot\exp\left(\frac{q\cdot \psi-0.5\cdot E_g}{k_BT}\right), 
	\label{eq:n_new}
\end{equation}
because the numerical challenges are now concentrated in one exponent, and $N_c\approx N_v$ are the regular effective density-of-states which scale as $\propto T^{3/2}$ and therefore are not challenging numerically. Eq. (\ref{eq:n_new}) solves (i), but (ii) and (iii) are still active. 

\subsubsection{Avoid Entering into the Numerically Forbidden Range}
\begin{figure}[t]
	\centering
	\includegraphics[width=0.35\textwidth]{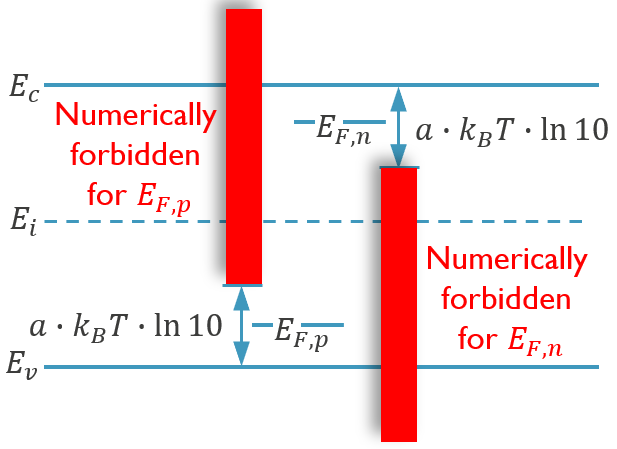}
	\caption{Numerically forbidden ranges in the bandgap for $E_{F,n}$ (electrons) and $E_{F,p}$ (holes). At \SI{300}{\kelvin}, the bandgap of most semiconductors is entirely numerically safe ($308\cdot k_BT\cdot \ln10 \approx\SI{18}{\electronvolt}$), but, at \SI{4.2}{\kelvin}, almost the entire bandgap will be numerically forbidden, except for small energy ranges close to each band edge ($308\cdot k_BT\cdot \ln10 \approx\SI{0.25}{\electronvolt}$). At \SI{1}{\micro\kelvin}, this range becomes exceptionally small ($\approx\SI{60}{\nano\electronvolt}$).}
	\label{fig:bandgap}
\end{figure}

To solve (ii), we must have that 
\begin{equation}
	-a\cdot\ln(10)\leqslant \frac{q\cdot \psi -0.5\cdot E_g}{k_BT}\leqslant a\cdot \ln(10)
	\label{eq:range}
\end{equation}
where $a<308$ to stay in double precision arithmetic. This is the same as preventing $E_{F,n}$ from straying too far from the conduction band edge (for $n$) and entering into the numerically forbidden range in the bandgap. This can be seen by using the definition of $\psi$ from Sec. \ref{sec:challenges} in (\ref{eq:range}), which shows that 
\begin{equation}
		E_{F,n}\geqslant E_c-a\cdot k_BT\ln(10),
\end{equation}
is required to avoid underflow. Similarly, for holes we would find that $E_{F,p}\leqslant E_v + a\cdot k_BT\ln(10)$. This creates the numerically forbidden ranges in the bandgap as shown in Fig. \ref{fig:bandgap}. 
These forbidden energy windows for $E_{F,n}$ and $E_{F,p}$ are better avoided, because they can translate into forbidden simulation domains within the diode (or any other device). Such division between \textquotedblleft allowed\textquotedblright \, and \textquotedblleft forbidden\textquotedblright \, simulation domains will be especially cumbersome because the boundary between them might change with material composition, temperature, device architecture, etc. Therefore it is important to avoid the underflow and overflow immediately in the distribution functions before they have a chance to ripple to other semiconductor quantities. To this end, we can enclose the Boltzmann exponent of (\ref{eq:n_new}) within a numerically safe sigmoid function $S(\eta,a)$, i.e.,
\begin{eqnarray}
	\label{eq:nsn_psi}n&=&N_c\cdot\exp\left[S\left(\frac{q\cdot \psi-0.5\cdot E_g}{k_BT},a\right)\right]
\end{eqnarray}
where 
\begin{equation}
	S(\eta,a)=a\cdot\ln(10)\cdot \tanh\left(\frac{\eta}{a\cdot \ln10}\right)
	\label{eq:Stanh}
\end{equation}
is the hyperbolic tangent function that goes from $-a\cdot\ln(10)$ to $a\cdot \ln(10)$, limiting the exponential in (\ref{eq:nsn_psi}) to $10^{-a}$ and $10^{a}$, respectively. A different sigmoid function could also be used, as long as it does not re-introduce troublesome exponentials, and preferably it should be an invertible function. Note that inserting $S$ is physically insignificant if the  precision parameter \textquotedblleft$a$\textquotedblright \, is chosen high enough (e.g., $a=200$), yet low enough to avoid numerical issues ($a<308$). For more details about $S$, and a semi-rigorous derivation, see \cite{beckers_bounded}.

\begin{figure}[t]
	\centering
	\includegraphics[width=0.46\textwidth]{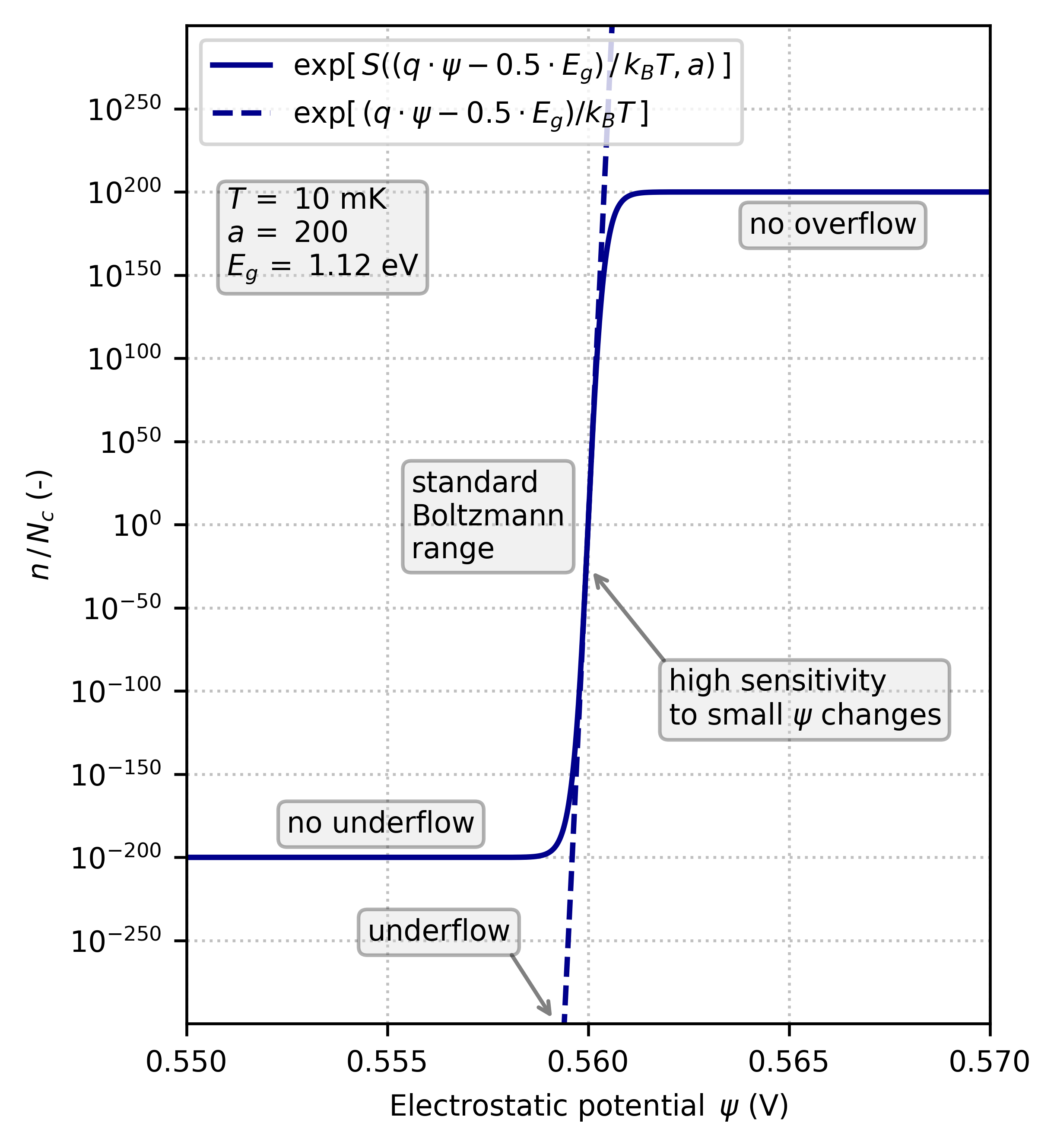}
	\caption{Exponentials from (\ref{eq:nsn_psi}) in logarithmic scale with no underflow nor overflow if $S$ is included, but still a high sensitivity to variations in $\psi$. This sensitivity is intrinsic to $T$, and must be tackled with a normalization to temperature of the solution variable in Poisson's equation.}
	\label{fig:eldens}
\end{figure}

Fig. \ref{fig:eldens} plots the exponentials from (\ref{eq:nsn_psi}) in logarithmic scale, at $T=\SI{10}{\milli\kelvin}$, with and without $S$. As can be seen in this figure, for physical levels of the carrier concentration, we have the regular Boltzmann exponential that is active, while underflow and overflow are avoided in the unphysical ranges. However, we must note that there is still a steep slope in the physical range, which is directly related to temperature, and therefore difficult to avoid, producing a large sensitivity to changes in $\psi$, i.e., the final problem (iii) must still be overcome. 
\subsubsection{\label{sec:sens}Avoid the Extreme Sensitivity by Normalizing}
To overcome (iii), the temperature dependences in $n$ and $p$ must be taken out of their exponents by solving (\ref{eq:poisson}) for a newly declared variable that is normalized to temperature, e.g., solve for the dimensionless $\eta\triangleq\left(q\cdot \psi-0.5\cdot E_g\right)/k_BT$ instead of $\psi$, which gives the following transformed Poisson's equation,
\begin{equation}
	\frac{\partial^2 \eta}{\partial x^2}=\frac{-\rho(\eta)}{\varepsilon_{si}},
	\label{eq:poisson_eta}
\end{equation}
where $\rho(\eta)=\frac{q^2}{k_BT}\times$
\begin{equation}
	\left(-N_c\cdot e^{S(\eta,a)}+N_v\cdot e^{-S\left(\eta+\frac{q\cdot E_g}{k_BT},a\right)}-N_A+N_D\right).\label{eq:rho_eta}
\end{equation}
Note that this normalization strategy of the electrostatic potential to temperature is only effective in combination with $S$. Also note that the remaining temperature dependence in the second exponent of (\ref{eq:rho_eta}) is not sensitive because there is no potential in the numerator anymore, only a fixed $E_g$. 

In a Newton-Raphson scheme, one also needs the derivative of the charge density, which is given by
\begin{equation}
	\frac{\partial \rho(\eta)}{\partial \eta}=\frac{-q^2}{k_BT}\cdot \left( N_c\cdot e^{S(\eta,a)} + N_v\cdot e^{-S\left(\eta+\frac{q\cdot E_g}{k_BT},a\right)}\right),
\end{equation}
where $S$ was ignored in the chain rule. 
\begin{figure*}[t]
	\centering
	\includegraphics[width=0.95\textwidth]{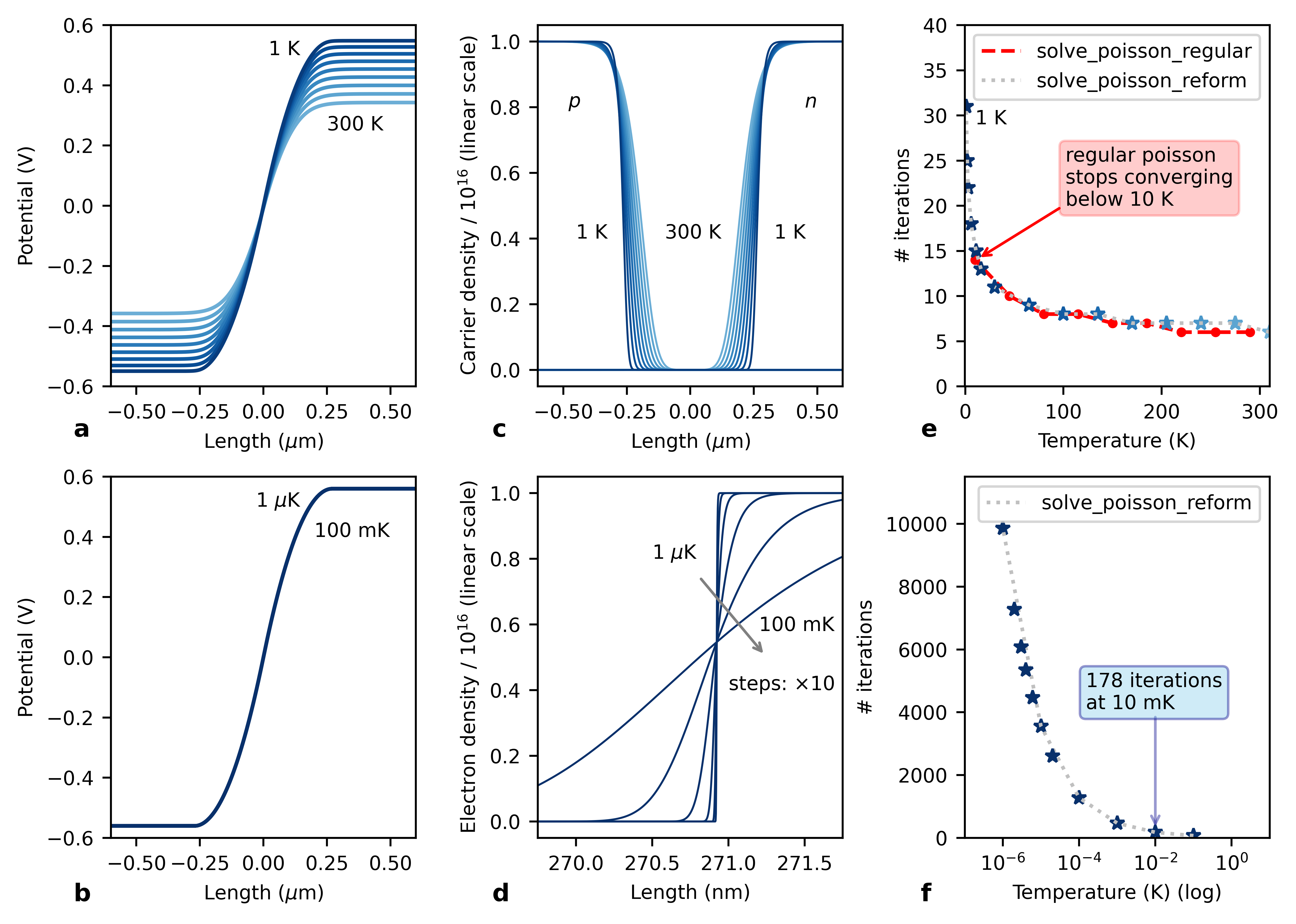}
	\caption{(a) Electrostatic potential from \SI{300}{\kelvin} down to \SI{1}{\kelvin} using \texttt{solve\symbol{95}poisson\symbol{95}reform} (double precision + reformulated Poisson), (b) Electrostatic potential from \SI{100}{\milli\kelvin} down to \SI{1}{\micro\kelvin} using \texttt{solve\symbol{95}poisson\symbol{95}reform}, (c) Electron and hole density from \SI{300}{\kelvin} down to \SI{1}{\kelvin} using \texttt{solve\symbol{95}poisson\symbol{95}reform}, (d) The $x$-axis zooms in on the electron density on the $n$-side to better resolve the rapid density variation around \SI{1}{\micro\kelvin}, (e) Number of Newton iterations required to reach convergence, and (f) Number of iterations increases rapidly below \SI{1}{\kelvin}.}
	\label{fig:slopes}
\end{figure*}
\subsubsection{Modify the Neumann Boundary Conditions} The standard boundary potentials, $\psi_p=(k_BT/q)\cdot \ln(n_i/N_A)$ and $\psi_n=(k_BT/q)\cdot \ln(N_D/n_i)$ in Fig.\ref{fig:detailspn}, need to be modified accordingly to suit the newly formulated Poisson's equation in terms of $\eta$ and including $S$. Using (\ref{eq:rho_eta}), charge neutrality on both sides of the junction imposes the following $\eta$'s at the edges :
\begin{eqnarray}
	\eta_n&=&a\cdot\ln(10)\cdot \atanh\left(\frac{\ln(N_D/N_c)}{a\cdot \ln10}\right)\label{eq:eta_n}\\
	\eta_p&=&\frac{-q\cdot E_g}{k_BT}+a\ln(10)\cdot \atanh\left(\!\frac{\ln(N_v/N_A)}{a\cdot \ln10}\!\right)\label{eq:eta_p}
\end{eqnarray}
which made use of the fact that $S$ is an invertible function. 

Besides these changes to Poisson's equation, a progressively finer mesh and weaker convergence criterion are also required to reach low-temperature convergence. Most of the proposed transformations cannot be implemented through user-defined functions in commercial TCAD; therefore a small demonstrator will be set up in Python in the next section. 

\section{\label{sec:appli3}Microkelvin Convergence Demonstration}
In this section, we compare the convergence of two Poisson solvers at low temperatures. The two Python functions implementing these Poisson solvers are presented in Appendix \ref{sec:app}. The first solver, \verb|solve_poisson_standard|, is the reference Poisson solver using double precision arithmetic without any changes to Poisson's equation (see e.g., \cite{jabr}). The second one, \verb|solve_poisson_reform|, implements all transformations that were discussed in Sec. \ref{sec:transf}, and also retains the double precision arithmetic. In both functions, Poisson's equation is discretized on a uniform grid (as shown in Fig.\ref{fig:detailspn}), cast into a system of non-linear difference algebraic equations, and then solved iteratively using Newton-Raphson. The minimum grid size is set by the Debye length $L_D=\sqrt{\varepsilon_{si} k_BT/(q^2N_A)}$ \cite{vasileska}, which reduces at lower temperatures, thus a sufficiently dense grid was used in all simulations ($m=75\,000$ grid points down to \SI{1}{\kelvin}, and $m=600\,000$ down to \SI{1}{\micro\kelvin}). Furthermore, we have used $E_g=\SI{1.12}{\electronvolt}$ and $a=100$. The Newton iteration is set to terminate when the error is less than $5.5\times 10^{-7}$. While the convergence criterion can easily reach machine precision at \SI{300}{\kelvin}, it seems difficult to converge with a criterion smaller than $5.5\times 10^{-7}$ at microkelvin temperatures. Even when including the bounded distribution function and variable transformation, sub-Kelvin simulation is still an ill-conditioned problem.

Fig. \ref{fig:slopes} presents the simulation results obtained in Python. Figs.\ref{fig:slopes}(a) and \ref{fig:slopes}(b) show the electrostatic potential diagram in the diode down to \SI{1}{\kelvin}, and \SI{1}{\micro\kelvin}, respectively, obtained using \verb|solve_poisson_reform|. The built-in potential increases at lower temperatures, but the variation becomes minimal below \SI{1}{\kelvin}. In Figs.\ref{fig:slopes}(c)-(d), we are able to resolve the rapid rise of the carrier densities at the edges of the depletion layer for the first time. Figs.\ref{fig:slopes}(e)-(f) show the number of Newton iterations required to reach convergence at each $T$. 

As expected from Fig. \ref{fig:ni}, \verb|solve_poisson_standard| indeed fails to converge below $\SI{10}{\kelvin}$ in double precision arithmetic [shown in red in Fig.\ref{fig:slopes}(e)]. On the other hand, thanks to the reformulation of Poisson's equation, \verb|solve_poisson_reform| succeeds in simulating down to \SI{1}{\micro\kelvin} (possibly lower), which is an improvement over several orders of magnitude as compared to the latest electrostatic diode simulation at \SI{30}{\kelvin} [Fig.2, \cite{kantner}]. Yet, despite these transformations, the number of required Newton iterations increases exponentially below \SI{50}{\kelvin}, although the required number is still very reasonable around $\approx\SI{10}{\milli\kelvin}$ (178 iterations), and $\approx$ \SI{4.2}{\kelvin} (about 20 iterations), which are currently the most relevant temperatures for experimental quantum science and the development of deep-cryogenic electronic circuits. 

\section{\label{sec:conclusions}Conclusions \& Outlook}
\begin{itemize}
	\item Microkelvin temperature convergence was demonstrated for the electrostatics in a one-dimensional $p\textendash n$ diode using the IEEE-754 double precision format. 
	\item Poisson's equation was solved iteratively down to \SI{1}{\micro\kelvin}, returning the potential and carrier densities without running into underflow, overflow, or convergence issues.  Nevertheless, we observed a significant increase in the required number of Newton iterations.
	\item To achieve this, step-by-step transformations were first presented for Poisson's equation to improve its numerical robustness and convergence. We discussed the forbidden energy ranges in the bandgap for the electron and hole Fermi levels. We suggested the use of a sigmoid function in the exponent to avoid entering these ranges.
	\item Steepness of Boltzmann's exponential tail is a numerically dangerous yet indispensable feature of temperature causing an extreme sensitivity to small variations in the potential. It is therefore recommended to solve Poisson's equation for a dimensionless variable that normalizes the potential to the thermal voltage. However, normalization is only effective in combination with the numerically safe sigmoid function in the exponent. 
	\item The proposed \SI{1}{\micro\kelvin} example written in Python can serve as a blueprint for further reducing the lowest achievable temperature in commercial TCAD and other device simulators, closing the gap with recent experimental progress.
\end{itemize}

\appendices
\section{\label{sec:app}Standard and Reformulated Poisson Solvers in Python Code for $p$ \textendash\,$n$ Diode}
The Python functions given below use the symbol \verb|V| for the electrostatic potential (instead of $\psi$ used in the main text) and \verb|E| for $\eta$ (normalized potential introduced in Section \ref{sec:sens}). 
\begin{lstlisting}[language=Python]
import numpy as np
from scipy.constants import e, k, m_e, h
from scipy import sparse
from scipy.sparse import linalg
\end{lstlisting}

\begin{lstlisting}[language=Python]
def solve_poisson_standard(T, NAp, NDn, Eg, L, m, number_eps):   # stops converging below 10 K
		eps_si = 1.05e-12	  # silicon permittivity
		UT = k * T / e      # thermal voltage
		theta = L / m       # discretization step
		
		ni = n_i(T, Eg) * 1e-6	# intrinsic carrier density
		
		V = np.zeros(m + 2)     # initial guess
		VN = UT * np.log(NDn / ni)
		VP = -UT * np.log(NAp / ni)
		V[0:int(m/2)+1] = VP
		V[int(m/2)+1:m+2] = VN
		
		NA = np.zeros(m+2)			# abrupt doping profile
		NA[0:int(m/2)+1] = NAp
		ND = np.zeros(m+2)
		ND[int(m/2)+1:m+2] = NDn
		
		eps = 2.2204e-16		   # machine precision
		Error = 1000000000000 * eps
		counter = 0
		while Error > number_eps * eps:# convergence criterion
					d2V_by_dx2 = (V[0:m] - 2 * V[1:m+1]                                       + V[2:m+2]) / theta**2
					rho = e * (ND[1:m+1] - NA[1:m+1]-ni*np.exp(V[1:m             +1] / UT) + ni * np.exp(-V[1:m+1] / UT))
					R = d2V_by_dx2 + rho / eps_si
					Mj = 2/theta**2 + (e * ni / (eps_si * UT))             	*(np.exp(V[1:m+1]/UT) + np.exp(-V[1:m+1]/UT))
					left_diag = (-1 / theta**2) * np.ones(m-1)
					right_diag = (-1 / theta**2) * np.ones(m-1)
					diags = np.array([left_diag, Mj, right_diag])
					CM = sparse.diags(diags, np.array([-1, 0, 1]),                     shape=(m, m))
					DV = linalg.spsolve(CM, R)
					V[1:m+1] = V[1:m+1] + DV   # update potential
					Error = np.linalg.norm(DV, 2) / np.sqrt(m)
				  counter = counter + 1
		return V[1:m + 1], counter, Error
\end{lstlisting}
\begin{lstlisting}[language=Python]
def solve_poisson_reform(T, NAp, NDn, Eg, L, m, number_eps, a):  # converges down to 1 uK (possibly lower) 
		eps_si = 1.05e-12		
		UT = k * T / e			
		theta = L / m		

		E = np.zeros(m+2) # modified Neumann conditions
		EN = a * np.log(10) * np.arctanh(np.log(NDn * 1e6 / return_Nc(T)) / (a * np.log(10)))  # Eqs. (13) & (14)
		EP = -Eg / UT + a * np.log(10) * np.arctanh(np.log(return_Nv(T) / (NAp * 1e6)) / (a * np.log(10)))  
		E[0:int(m/2)+1] = EP
		E[int(m/2)+1:m+2] = EN

		NA = np.zeros(m+2)
		NA[0:int(m/2)+1] = NAp
		ND = np.zeros(m+2)
		ND[int(m/2)+1:m+2] = NDn

		Nc = return_Nc(T) * 1e-6  # use Nc & Nv instead of ni
		Nv = return_Nv(T) * 1e-6

		eps = 2.2204e-16
		Error = 1000000000000 * eps
		counter = 0
		while Error > number_eps * eps:
					d2E_by_dx2 = (E[0:m] - 2 * E[1:m+1]                                       + E[2:m+2]) / theta**2
					# reformulated Poisson equation including S
					rho = (e / UT) * (ND[1:m+1] - NA[1:m+1]                    - Nc * np.exp(S(E[1:m+1], a))                        + Nv * np.exp(-S(E[1:m+1] + Eg/UT, a))) 
					R = d2E_by_dx2 + rho / eps_si
					Mj = 2/theta**2 + (e / (eps_si * UT)) *                   (Nc * np.exp(S(E[1:m+1], a))                          + Nv * np.exp(-S(E[1:m+1] + Eg/UT, a)))
					left_diag = (-1 / theta**2) * np.ones(m-1)
					right_diag = (-1 / theta**2) * np.ones(m-1)
					diags = np.array([left_diag, Mj, right_diag])
					CM = sparse.diags(diags, np.array([-1, 0, 1]),                           shape=(m, m))
					# solve for normalized potential 
					DE = linalg.spsolve(CM, R) 
					E[1:m+1] = E[1:m+1] + DE
				  Error = np.linalg.norm(DE, 2) / np.sqrt(m)
				  counter = counter + 1
		return E[1:m+1], counter, Error
\end{lstlisting}		

Auxiliary functions:
\begin{lstlisting}[language=Python]
def return_Nc(t):# effective conduction band DOS [#/(m^3)]
		return 2*((2 * np.pi * 1.182 * m_e * k * t) / (h**2))**(3/2)
	
def return_Nv(t):# effective valence band DOS [#/(m^3)]
		return 2*((2 * np.pi * 0.81 * m_e * k * t) / (h**2))**(3/2)
	
def S(eta, a):   # numerically safe sigmoid function
		return a*np.log(10) * np.tanh(eta / (a * np.log(10)))
		
def n_i(t, Eg):  # intrinsic carrier density [#/(m^3)]
		return np.sqrt(return_Nc(t) * return_Nv(t)) * np.exp(-Eg * e / (2*k*t))
\end{lstlisting}

\bibliographystyle{ieeetran}
\bibliography{citations}

\begin{thebibliography}{10}
\providecommand{\url}[1]{#1}
\csname url@samestyle\endcsname
\providecommand{\newblock}{\relax}
\providecommand{\bibinfo}[2]{#2}
\providecommand{\BIBentrySTDinterwordspacing}{\spaceskip=0pt\relax}
\providecommand{\BIBentryALTinterwordstretchfactor}{4}
\providecommand{\BIBentryALTinterwordspacing}{\spaceskip=\fontdimen2\font plus
\BIBentryALTinterwordstretchfactor\fontdimen3\font minus
  \fontdimen4\font\relax}
\providecommand{\BIBforeignlanguage}[2]{{%
\expandafter\ifx\csname l@#1\endcsname\relax
\typeout{** WARNING: IEEEtran.bst: No hyphenation pattern has been}%
\typeout{** loaded for the language `#1'. Using the pattern for}%
\typeout{** the default language instead.}%
\else
\language=\csname l@#1\endcsname
\fi
#2}}
\providecommand{\BIBdecl}{\relax}
\BIBdecl

\bibitem{selberherr}
S.~Selberherr, ``{MOS} device modeling at 77 {K},'' \emph{IEEE Transactions on
  Electron Devices}, vol.~36, no.~8, pp. 1464--1474, Aug. 1989,
  \href{https://ieeexplore.ieee.org/document/30960}{doi:10.1109/16.30960}.

\bibitem{kantner}
M.~Kantner and T.~Koprucki, ``Numerical simulation of carrier transport in
  semiconductor devices at cryogenic temperatures,'' \emph{Optical and Quantum
  Electronics}, vol.~48, no.~12, Dec. 2016,
  \href{http://link.springer.com/10.1007/s11082-016-0817-2}{doi:10.1007/s11082-016-0817-2}.

\bibitem{gao}
X.~Gao, E.~Nielsen, R.~P. Muller, R.~W. Young, A.~G. Salinger, N.~C. Bishop,
  M.~P. Lilly, and M.~S. Carroll, ``Quantum computer aided design simulation
  and optimization of semiconductor quantum dots,'' \emph{Journal of Applied
  Physics}, vol. 114, no.~16, p. 164302, Oct. 2013, doi:
  \href{http://aip.scitation.org/doi/10.1063/1.4825209}{10.1063/1.4825209}.

\bibitem{mohiyaddin_multiphysics_2019}
F.~A. Mohiyaddin, B.~Chan, T.~Ivanov, A.~Spessot, P.~Matagne, J.~Lee,
  B.~Govoreanu, I.~P. Radu, G.~Simion, N.~I.~D. Stuyck, R.~Li, F.~Ciubotaru,
  G.~Eneman, F.~M. Bufler, S.~Kubicek, and J.~Jussot, ``Multiphysics
  {Simulation} \& {Design} of {Silicon} {Quantum} {Dot} {Qubit}
  {Devices}.''\hskip 1em plus 0.5em minus 0.4em\relax IEEE, Dec. 2019, pp.
  39.5.1--39.5.4,
  \href{https://ieeexplore.ieee.org/document/8993541/}{10.1109/IEDM19573.2019.8993541}.

\bibitem{richey}
\BIBentryALTinterwordspacing
D.~M. Richey, J.~D. Cressler, and R.~C. Jaeger, ``Numerical simulation of
  {SiGe} {HBT}'s at cryogenic temperatures,'' \emph{Le Journal de Physique IV},
  vol.~04, pp. C6--127--C6--132, Jun. 1994. [Online]. Available:
  \url{http://www.edpsciences.org/10.1051/jp4:1994620}
\BIBentrySTDinterwordspacing

\bibitem{tedpaper}
A.~Beckers, F.~Jazaeri, and C.~Enz, ``Cryogenic {MOS} {Transistor} {Model},''
  \emph{IEEE Transactions on Electron Devices}, vol.~65, no.~9, pp. 3617--3625,
  Sep. 2018,
  \href{https://ieeexplore.ieee.org/document/8424046/}{doi:10.1109/TED.2018.2854701}.

\bibitem{zlatan}
\BIBentryALTinterwordspacing
Z.~Stanojevic, J.~M. Gonzalez~Medina, F.~Schanovsky, and M.~Karner,
  ``{Quasi-Fermi-Based Charge Transport Scheme for Device Simulation in
  Cryogenic, Wide-Band-Gap, and High-Voltage Applications},'' \emph{{Preprint
  Submitted to Transactions on Electron Devices}}. [Online]. Available:
  \url{https://doi.org/10.36227/techrxiv.21132637.v1}
\BIBentrySTDinterwordspacing

\bibitem{catapano}
E.~Catapano, M.~Cassé, F.~Gaillard, S.~{de Franceschi}, T.~Meunier, M.~Vinet,
  and G.~Ghibaudo, ``{TCAD} {Simulations} of {FDSOI} devices down to {Deep}
  {Cryogenic} {Temperature},'' \emph{Solid-State Electronics}, p. 108319, 2022,
  doi:
  \href{https://www.sciencedirect.com/science/article/pii/S0038110122000910}{10.1016/j.sse.2022.108319}.

\bibitem{aouad}
M.~Aouad, T.~Poiroux, S.~Martinie, F.~Triozon, M.~Vinet, and G.~Ghibaudo,
  ``Poisson-{Schrödinger} simulation and analytical modeling of inversion
  charge in {FDSOI} {MOSFET} down to 0 {K} – {Towards} compact modeling for
  cryo {CMOS} application,'' \emph{Solid-State Electronics}, vol. 186, p.
  108126, Dec. 2021,
  \href{https://linkinghub.elsevier.com/retrieve/pii/S0038110121001696}{doi:10.1016/j.sse.2021.108126}.

\bibitem{beaudoin_robust_2022}
\BIBentryALTinterwordspacing
F.~Beaudoin, P.~Philippopoulos, C.~Zhou, I.~Kriekouki, M.~Pioro-Ladrière,
  H.~Guo, and P.~Galy, ``Robust technology computer-aided design of gated
  quantum dots at cryogenic temperature,'' \emph{Applied Physics Letters}, vol.
  120, no.~26, p. 264001, Jun. 2022. [Online]. Available:
  \url{https://aip.scitation.org/doi/10.1063/5.0097202}
\BIBentrySTDinterwordspacing

\bibitem{moroz}
\BIBentryALTinterwordspacing
V.~Moroz, J.~Kawa, X.-W. Lin, A.~R. Brown, P.~Asenov, J.~Lee, M.~Bajaj,
  T.~Michalak, C.~Riddet, A.~Svizhenko, R.~Hentschke, and S.~Smidstrup,
  ``\BIBforeignlanguage{en}{Challenges in {Design} and {Modeling} of {Cold}
  {CMOS} {HPC} {Technology}},'' in \emph{\BIBforeignlanguage{en}{2021
  {International} {Conference} on {Simulation} of {Semiconductor} {Processes}
  and {Devices} ({SISPAD})}}.\hskip 1em plus 0.5em minus 0.4em\relax Dallas,
  TX, USA: IEEE, Sep. 2021, pp. 107--110. [Online]. Available:
  \url{https://ieeexplore.ieee.org/document/9592537/}
\BIBentrySTDinterwordspacing

\bibitem{microkelvin}
\BIBentryALTinterwordspacing
G.~Pickett and C.~Enss, ``\BIBforeignlanguage{en}{The {European} {Microkelvin}
  {Platform}},'' \emph{\BIBforeignlanguage{en}{Nature Reviews Materials}},
  vol.~3, no.~3, Mar. 2018. [Online]. Available:
  \url{http://www.nature.com/articles/natrevmats201812}
\BIBentrySTDinterwordspacing

\bibitem{samani_microkelvin_2022}
\BIBentryALTinterwordspacing
M.~Samani, C.~P. Scheller, O.~S. Sedeh, D.~M. Zumbühl, N.~Yurttagül,
  K.~Grigoras, D.~Gunnarsson, M.~Prunnila, A.~T. Jones, J.~R. Prance, and R.~P.
  Haley, ``\BIBforeignlanguage{en}{Microkelvin electronics on a pulse-tube
  cryostat with a gate {Coulomb}-blockade thermometer},''
  \emph{\BIBforeignlanguage{en}{Physical Review Research}}, vol.~4, no.~3, p.
  033225, Sep. 2022. [Online]. Available:
  \url{https://link.aps.org/doi/10.1103/PhysRevResearch.4.033225}
\BIBentrySTDinterwordspacing

\bibitem{kao}
\BIBentryALTinterwordspacing
K.-H. Kao, C.~Godfrin, A.~Elsayed, R.~Li, E.~Simoen, A.~Grill, S.~Kubicek,
  I.~P. Radu, and B.~Govoreanu, ``Linking {Room}- and {Low}-{Temperature}
  {Electrical} {Performance} of {MOS} {Gate} {Stacks} for {Cryogenic}
  {Applications},'' \emph{IEEE Electron Device Letters}, vol.~43, no.~5, pp.
  674--677, May 2022. [Online]. Available:
  \url{https://ieeexplore.ieee.org/document/9743910/}
\BIBentrySTDinterwordspacing

\bibitem{hart}
\BIBentryALTinterwordspacing
P.~A. 't~Hart, T.~Huizinga, M.~Babaie, A.~Vladimirescu, and F.~Sebastiano,
  ``Integrated {Cryo}-{CMOS} {Temperature} {Sensors} for {Quantum} {Control}
  {ICs},'' in \emph{2022 {IEEE} 15th {Workshop} on {Low} {Temperature}
  {Electronics} ({WOLTE})}.\hskip 1em plus 0.5em minus 0.4em\relax Matera,
  Italy: IEEE, Jun. 2022, pp. 1--4. [Online]. Available:
  \url{https://ieeexplore.ieee.org/document/9882600/}
\BIBentrySTDinterwordspacing

\bibitem{beckers_bounded}
\BIBentryALTinterwordspacing
A.~Beckers, ``{Bounded Distribution Functions for Applied Physics, Especially
  Electron Device Simulation at Deep-Cryogenic Temperatures},'' Dec. 2022.
  [Online]. Available: \url{https://arxiv.org/abs/2212.01786}
\BIBentrySTDinterwordspacing

\bibitem{jabr}
\BIBentryALTinterwordspacing
R.~A. Jabr, M.~Hamad, and Y.~M. Mohanna, ``Newton-{Raphson} {Solution} of
  {Poisson}'s {Equation} in a \textit{{Pn}} {Diode},'' \emph{The International
  Journal of Electrical Engineering \& Education}, vol.~44, no.~1, pp. 23--33,
  Jan. 2007. [Online]. Available:
  \url{http://journals.sagepub.com/doi/10.7227/IJEEE.44.1.3}
\BIBentrySTDinterwordspacing

\bibitem{vasileska}
\BIBentryALTinterwordspacing
D.~Vasileska, S.~M.~Goodnick, and G.~Klimeck, \emph{Computational
  {Electronics}: {Semiclassical} and {Quantum} {Device} {Modeling} and
  {Simulation}}, 1st~ed.\hskip 1em plus 0.5em minus 0.4em\relax CRC Press, Dec.
  2017. [Online]. Available:
  \url{https://www.taylorfrancis.com/books/9781420064841}
\BIBentrySTDinterwordspacing

\end{thebibliography}

\end{document}